\documentclass[preprint, 12pt]{elsarticle}
\usepackage[T1]{fontenc}    %
\usepackage[margin=1in]{geometry} %
\journal{}
\date{}

\usepackage{comment}
\usepackage{ulem}

\usepackage{fancyhdr}

\usepackage{enumitem}

\usepackage{amsmath,amsthm,amsfonts,amssymb,amscd}
\usepackage{upgreek} %
\usepackage{mathrsfs} %
\usepackage{mathtools} %
\usepackage{derivative}

\usepackage{thmtools,thm-restate}
\usepackage{xcolor}

\usepackage{float}
\usepackage{placeins} %

\usepackage{graphicx}
\usepackage{tikz,tikz-cd}
\usetikzlibrary{arrows, shapes}
\usetikzlibrary{mindmap}
\tikzcdset{scale cd/.style={every label/.append style={scale=#1}, cells={nodes={scale=#1}}}}
\tikzcdset{trapezium stretches=true}
\tikzcdset{arrow style=tikz, diagrams={>=stealth'}}

\usepackage{booktabs}
\usepackage{multirow}
\usepackage{makecell}

\usepackage{caption}
\usepackage{subcaption}

\usepackage[linesnumbered,ruled]{algorithm2e}
\usepackage{listings}

\usepackage{tcolorbox}

\usepackage[colorlinks=true, citecolor=blue, filecolor=black, linkcolor=blue, urlcolor=gray]{hyperref}

\usepackage{natbib}

\usepackage{lastpage}

\usepackage[capitalise,nameinlink,noabbrev]{cleveref}
\crefname{section}{Sec.}{Sec.}
\Crefname{section}{Section}{Sections}
\crefname{subsection}{Sec.}{Sec.}
\Crefname{subsection}{Section}{Sections}
\crefname{figure}{Fig.}{Fig.}
\Crefname{figure}{Figure}{Figures}
\crefname{table}{Table}{Tables}
\Crefname{table}{Table}{Tables}
\crefname{equation}{Eq.}{Eq.}
\Crefname{equation}{Equation}{Equations}

\makeatletter

\begin{document}
\begin{frontmatter}
\title{Learning Population-Level Dynamics through a Latent Fokker--Planck Model and Discrepancy Transport Maps}

\author[USC]{Chengyang Huang}
\author[USC]{Krishna Garikipati}

\affiliation[USC]{
    organization={Department of Aerospace and Mechanical Engineering, University of Southern California},
    city={Los Angeles},
    postcode={CA 90089},
    country={United States}
}

\begin{abstract}
Many scientific and engineering systems are observed as time-indexed probability distributions whose governing dynamics are unknown and whose individual trajectories are unavailable. These settings challenge conventional system-identification approaches that rely on trajectory correspondence or prescribed evolution equations. This work presents a population-level inference framework that recovers latent stochastic dynamics directly from snapshot probability distributions by decomposing the observed evolution into an intrinsic latent stochastic process and a discrepancy transport map that captures geometric deformation between the latent and observed probability spaces. The latent dynamics are modeled using an Ornstein--Uhlenbeck process, providing a closed-form solution to the associated Fokker--Planck equation, while the discrepancy transport map is parameterized through the Knothe--Rosenblatt rearrangement with monotone neural networks. To mitigate the non-uniqueness inherent in the latent--transport decomposition, the transport map is regularized using a deformation energy motivated by hyperelasticity, promoting smooth, physically interpretable deformations while reducing unnecessary complexity. The latent stochastic model and discrepancy transport map are learned jointly through a unified optimization problem defined over probability distributions. Numerical examples involving nonlinear and multimodal distributional dynamics demonstrate that the proposed framework accurately reconstructs complex probability evolution while preserving a compact and analytically tractable latent representation. The proposed formulation provides a general framework for population-level dynamical inference and establishes a foundation for extending latent stochastic models and transport-based learning to more general and higher-dimensional systems.
\end{abstract}

\end{frontmatter}

\section{Introduction}
Many scientific and engineering systems are naturally observed through the evolution of population-level probability distributions rather than trajectories of individual entities.
Representative examples include particle systems, molecular dynamics, diffusion processes, and microstructural evolution in materials~\cite{risken1996fokker,Frenkel2001-rd,Pavliotis2014}, as well as single-cell transcriptomics and flow cytometry in computational biology~\cite{Schiebinger2019,Bendall2011}. Despite their diverse physical origins, these applications share a common observation model: at each observation time, only a population snapshot is available, represented by a collection of independent samples drawn from an underlying probability distribution.

This observation setting introduces two fundamental challenges.
First, the time-dependence of probability distributions in the observation space often do not conform to canonical dynamical descriptions, hence masking the intrinsic stochastic evolution of the underlying system. Secondly, the absence of temporal correspondence fundamentally changes the inference problem. Instead of reconstructing individual trajectories, the objective is to infer the underlying stochastic process directly from a sequence of snapshot distributions.
Observation processes, nonlinear measurement operators, geometric distortions, and experimental noise may substantially alter the probability distributions observed in the physical space.
Consequently, the observed probability evolution generally reflects both the intrinsic stochastic evolution of the underlying system in a latent space and its transformation to the observation space.
The objective is therefore to recover a decomposition that distinguishes these two components directly from the observed probability distributions.

Most existing learning methods for dynamical systems assume access to trajectory information. State-space models, Kalman filtering and smoothing methods, neural ordinary differential equations, latent stochastic differential equations (SDEs), and related latent-variable models infer hidden dynamics by exploiting temporal correspondence between successive observations~\cite{Kalman1960,Rauch1965,krishnan2015deepkalmanfilters,Chen2018NeuralODE,Rubanova2019LatentODE,Li2020LatentSDE,Morton2019DeepKoopman}. Their primary object of inference is the evolution of latent trajectories. When only snapshot observations are available, however, this temporal information is absent, making these approaches difficult to apply directly.

When trajectory information is unavailable, the natural object of inference becomes the probability distribution itself rather than individual realizations of the stochastic process. This observation has motivated a broad class of methods that directly model the evolution of probability distributions.
Examples include Fokker--Planck (FP) equations for stochastic dynamical systems, optimal transport formulations for probability evolution, and more recently transport-map-based methods that represent complex probability distributions through invertible transformations of simpler reference distributions~\cite{risken1996fokker,Pavliotis2014,Villani2009,Santambrogio2015,Brenier1991,Baptista2023ATM}.
Although these approaches successfully describe the evolution of observed probability distributions, they generally treat the observed distributions as the primary modeling target.
However, this perspective may fail if the observed dynamics do not align with the canonical form of a description, such as the FP equation. Put differently, these methods do not explicitly distinguish between the intrinsic stochastic evolution of the underlying system and the transformations introduced by the observation process.

To address these challenges, we formulate population-level dynamical inference as an inverse problem over evolving probability measures. We seek a decomposition of the observed probability evolution into two complementary components: an intrinsic latent probability measure governed by stochastic dynamics and a transport map relating the latent and observation spaces. The latent stochastic evolution captures the intrinsic stochastic behavior of the underlying system, whereas the transport map accounts for observation-induced geometric transformations. Because the latter diffeomorphic map bridges any mismatch between the canonical form of the stochastic dynamics in the latent space and its non-conforming appearance in observation space, we refer to it as a discrepancy transport map.

The proposed latent--transport decomposition is generally non-unique, since multiple latent evolutions and discrepancy transport maps may reproduce the same observed probability evolution.
Consequently, the objective is not merely to identify one admissible decomposition, but rather to identify the decomposition that  obeys a minimum principle even while explaining the observations.
This perspective closely parallels kinematic decompositions in continuum mechanics. The direct example comes from the coupling of mass transport and elasticity in two- or three-dimensional elastic solids: The mass of some tracer material undergoes transport in an undeformed reference configuration  simultaneously with finite deformation of the elastic solid. The deformation, generally a diffeomorphism, is subject to the energy minimization of hyperelasticity. It further transports the tracer mass leading to the observed dynamics, which however, may not subscribe to a canonical form such as the FP equation. A further example is finite-strain plasticity, which interprets the observed deformation through a multiplicative decomposition into elastic and plastic components, neither of which is directly observable~\cite{holzapfel2002nonlinear,bigoni2012nonlinear}. However, the elastic component parametrizes the elastic energy, which enters the minimization principle. The plastic component evolves under the guidance of other microstructural processes. Likewise, we interpret the observed probability evolution as the combined effect of an intrinsic stochastic evolution and transport to the observed space. As in the cases from continuum mechanics, the decomposition cannot be uniquely determined from observations alone and therefore requires additional physical principles to identify a meaningful solution.
Motivated by this analogy, we regularize the latent--transport decomposition through physically motivated principles: the latent evolution is constrained by an FP equation, while the discrepancy transport map is regularized by an elasticity-inspired deformation energy.

The main contributions of this work are summarized as follows:

\begin{itemize}

\item We propose a latent--transport decomposition in which the latent stochastic evolution is constrained by an FP equation while the observed probability evolution is represented through an invertible transport map.

\item We formulate population-level dynamical inference as an inverse problem over evolving probability measures using only snapshot observations, without requiring trajectory correspondence.

\item We develop a unified inference framework that jointly estimates latent stochastic dynamics and discrepancy transport maps while regularizing the non-uniqueness of the latent--transport decomposition through elasticity-inspired transport regularization.

\end{itemize}

The remainder of this paper is organized as follows.
\Cref{sec:methodology} presents the proposed framework, including the problem formulation, the latent stochastic evolution governed by an FP equation, the transport-map representation, and the joint inference algorithm.
\Cref{sec:numerical_results} demonstrates the proposed method on two synthetic examples designed to validate different aspects of the latent--transport decomposition.
\Cref{sec:discussion} discusses the interpretation of the proposed decomposition, its relationship to existing approaches, its limitations, and directions for future research.
Finally, \Cref{sec:conclusion} concludes the paper.

\section{Methodology}
\label{sec:methodology}
In developing the proposed framework, we first formulate the inference problem, then introduce the latent stochastic evolution and discrepancy transport-map representation, followed by the joint estimation framework.

\subsection{Problem formulation}
\label{subsec:problem}

Let $\mathcal{X}\subset\mathbb{R}^{d}$ denote the observation space, with coordinate $\boldsymbol{x}\in\mathcal{X}$, and let $\mathcal{T}\subset\mathbb{R}_{\ge0}$ denote the time domain.
At each observation time $t\in\mathcal{T}$, the observed system is characterized by an unknown probability measure
$
\rho_t \in \mathcal{P}(\mathcal{X}),
$
where $\mathcal{P}(\mathcal{X})$ denotes the space of probability measures on $\mathcal{X}$. When $\rho_t$ admits a density with respect to Lebesgue measure, that density is denoted by $\rho(\boldsymbol{x},t)$. Throughout this paper, time-dependent functions are written with time as an explicit argument, for example $S(\boldsymbol{x},t)$ rather than $S_t(\boldsymbol{x})$; subscripts such as $\rho_t$ and $\zeta_t$ are reserved for time-indexed probability measures. The evolving distribution is observed only through a collection of independent and identically distributed (i.i.d.) samples,
\begin{equation}
\mathcal{D}
=
\left\{
\left\{
\boldsymbol{x}_{t_k}^{(i)}
\right\}_{i=1}^{m_k}
\right\}_{k=0}^{n},
\qquad
\boldsymbol{x}_{t_k}^{(i)}
\overset{\mathrm{i.i.d.}}{\sim}
\rho_{t_k},
\label{eq:snapshot_data}
\end{equation}
where $n+1$ is the total number of observation times and $m_k$ is the number of samples collected at observation time $t_k$. Temporal correspondence between samples collected at different observation times is unavailable. Accordingly, inference is performed directly from the observed snapshot distributions rather than individual trajectories.

\begin{figure}[hbtp]
    \centering
    \includegraphics[width=0.92\textwidth]{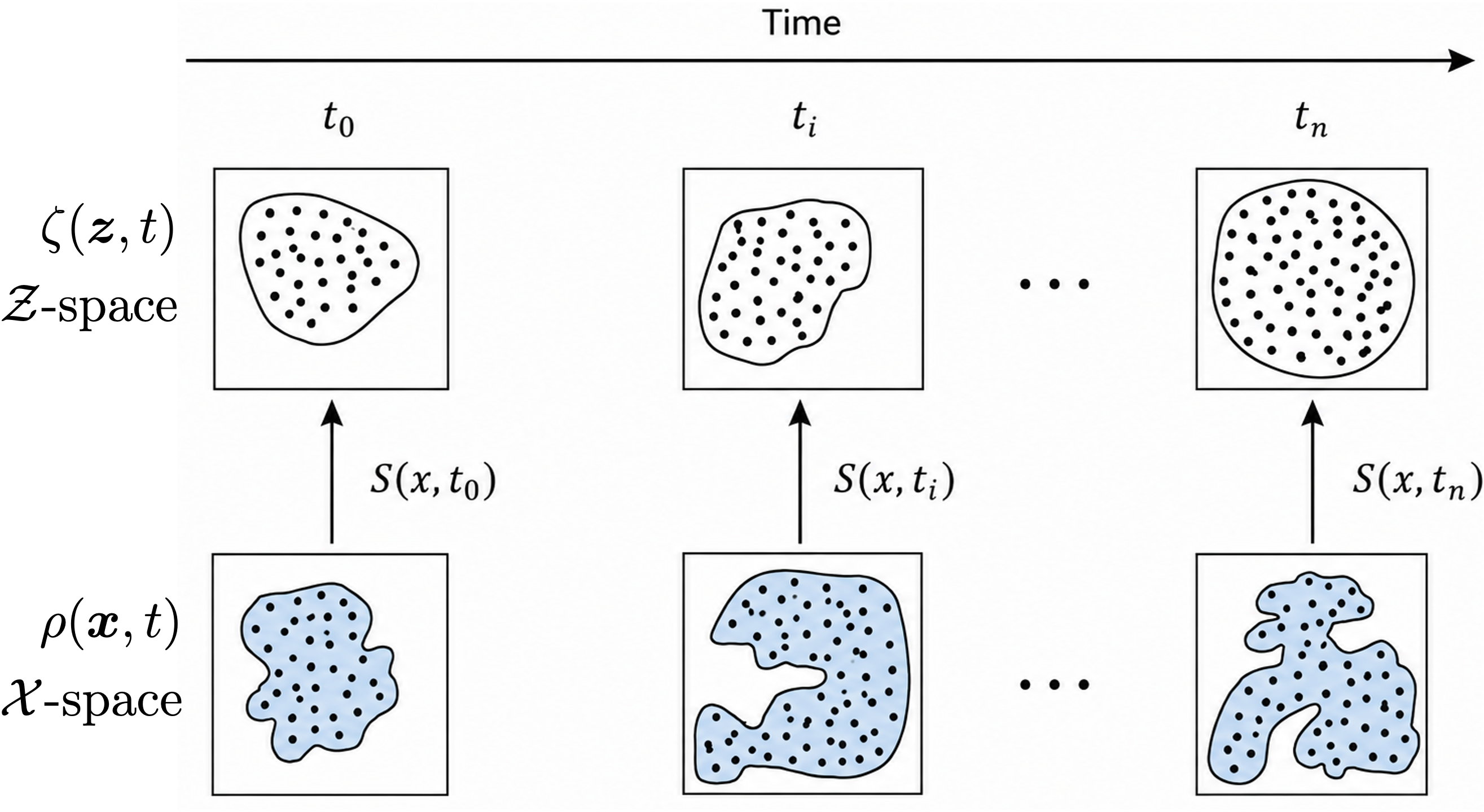}
    \caption{Conceptual illustration of the proposed latent--transport decomposition. The observed probability evolution is represented as the pushforward of a latent stochastic evolution through a time-dependent transport map. The latent stochastic evolution captures the intrinsic stochastic evolution, whereas the transport map represents the observation-induced geometric transformation. Both components are inferred jointly from snapshot observations without temporal correspondence.}
    \label{fig:problem_formulation}
\end{figure}

To model the latent--transport decomposition, we introduce a latent probability space $\mathcal{Z}\subset\mathbb{R}^{d}$, with coordinate $\boldsymbol{z}\in\mathcal{Z}$, and assume that, for each $t\in\mathcal{T}$, there exists a latent probability measure
$
\zeta_t\in\mathcal P(\mathcal{Z}),
$
representing the intrinsic stochastic evolution of the underlying system.
When $\zeta_t$ admits a density with respect to Lebesgue measure, that density is denoted by $\zeta(\boldsymbol{z},t)$.
We further introduce a time-dependent invertible transport map
$
T(\cdot,t):\mathcal{Z}\rightarrow\mathcal{X},
$
which models the geometric transformation between the latent and observation spaces. The observed probability measure is then represented as the pushforward of the latent probability measure,
\begin{equation}
\rho_t
=
T_{\sharp}(\cdot,t)\zeta_t,
\qquad
\forall\, t\in\mathcal T,
\label{eq:pushforward}
\end{equation}
where $T_{\sharp}(\cdot,t)$ denotes the pushforward operator induced by $T(\cdot,t)$.
The proposed decomposition is illustrated schematically in \cref{fig:problem_formulation}. The central objective is therefore to infer both the latent stochastic evolution and the transport map jointly from the observed snapshot distributions.

The decomposition is generally non-unique. In particular, let
$
G(\cdot,t):\mathcal{Z}\rightarrow\mathcal{Z}
$
be any invertible, time-dependent transformation of the latent space. Define the transformed latent probability measure and transport map by
\begin{equation}
\tilde{\zeta}_t
=
G(\cdot,t)_{\sharp}\zeta_t,
\qquad
\tilde{T}(\cdot,t)
=
T(\cdot,t)\circ G(\cdot,t)^{-1}.
\label{eq:reparameterization}
\end{equation}
Then
\begin{equation}
\tilde{T}(\cdot,t)_{\sharp}\tilde{\zeta}_t
=
T_{\sharp}(\cdot,t)\zeta_t
=
\rho_t,
\label{eq:equivalent_decomposition}
\end{equation}
showing that infinitely many latent--transport pairs generate the same observed probability evolution. Consequently, the latent stochastic evolution and transport map cannot be identified from the observations alone. Additional physical and mathematical constraints are therefore required to select a physically meaningful decomposition. In the following subsections, we constrain the latent stochastic evolution through an FP equation and regularize the transport map using an elasticity-inspired deformation energy.

\subsection{Latent dynamics governed by the Ornstein--Uhlenbeck process}
\label{subsec:latent_dynamics}

The latent stochastic evolution is modeled by an FP equation. Rather than allowing an arbitrary time-dependent evolution, we constrain the latent probability measure to satisfy a standard stochastic process with tractable mathematical form. This reduces the degrees of freedom of the inverse problem while ensuring that the latent evolution remains consistent with an underlying stochastic dynamics.
Specifically, the latent probability density satisfies
\begin{equation}
\frac{\partial \zeta(\boldsymbol{z},t)}{\partial t}
=
-\nabla\cdot
\left(
\boldsymbol{v}(\boldsymbol{z})\zeta(\boldsymbol{z},t)
\right)
+
\frac12
\nabla\cdot
\left(
\boldsymbol{D}_{\boldsymbol{z}}\nabla\zeta(\boldsymbol{z},t)
\right),
\label{eq:fp_general}
\end{equation}
where
$
\boldsymbol{v}:\mathcal{Z}\rightarrow\mathbb{R}^{d}
$
denotes the drift field and
$
\boldsymbol{D}_{\boldsymbol{z}}\in\mathbb{R}^{d\times d}
$
is a symmetric positive semidefinite diffusion tensor, and all spatial differential operators act with respect to the latent coordinate $\boldsymbol{z}$.

Among the stochastic processes governed by FP equations, we adopt the Ornstein--Uhlenbeck (OU) process as the latent stochastic dynamics. The OU process is the simplest nontrivial diffusion process whose associated FP equation admits a finite-dimensional closed-form evolution in which the transition density has a Gaussian form. Consequently, the latent stochastic evolution can be evaluated analytically throughout the observation interval without numerically solving the governing partial differential equation.
For the OU process, the drift field is the negative gradient of a quadratic potential $\psi(\boldsymbol{z})$, and is therefore affine,
\begin{equation}
\boldsymbol{v}(\boldsymbol{z}) = -\nabla\psi(\boldsymbol{z})
=
\boldsymbol{A}\boldsymbol{z}
+
\boldsymbol{b},
\label{eq:ou_drift}
\end{equation}
where the latent stochastic dynamics are parameterized by
\begin{equation}
\boldsymbol{\theta}_{\mathrm{OU}}
=
\left\{
\boldsymbol{A},
\boldsymbol{b},
\boldsymbol{D}_{\boldsymbol{z}}
\right\},
\end{equation}
which constitute the unknown parameters to be inferred.

The Gaussian family is invariant under the OU dynamics. If the initial latent probability measure is Gaussian,
\begin{equation}
\zeta(\boldsymbol{z},t_0)
=
\mathcal{N}
\left(
\boldsymbol{\mu}_0,
\boldsymbol{\Sigma}_0
\right),
\label{eq:gaussian_initial}
\end{equation}
then the latent probability density remains Gaussian for every
$t\in\mathcal{T}$,
\begin{equation}
\zeta(\boldsymbol{z},t)
=
\mathcal{N}
\left(
\boldsymbol{\mu}(t),
\boldsymbol{\Sigma}(t)
\right),
\label{eq:gaussian_family}
\end{equation}
whose evolution is completely characterized by the mean and covariance~\cite{risken1996fokker,Pavliotis2014}:
\begin{align}
\boldsymbol{\mu}(t)
&=
\exp
\!\left(
\boldsymbol{A}(t-t_0)
\right)
\boldsymbol{\mu}_0
+
\int_{t_0}^{t}
\exp
\!\left(
\boldsymbol{A}(t-s)
\right)
\boldsymbol{b}
\,\mathrm{d}s,
\label{eq:mean_solution}
\\
\boldsymbol{\Sigma}(t)
&=
\exp
\!\left(
\boldsymbol{A}(t-t_0)
\right)
\boldsymbol{\Sigma}_0
\exp
\!\left(
\boldsymbol{A}^{\mathsf T}(t-t_0)
\right)
+
\int_{t_0}^{t}
\exp
\!\left(
\boldsymbol{A}(t-s)
\right)
\boldsymbol{D}_{\boldsymbol{z}}
\exp
\!\left(
\boldsymbol{A}^{\mathsf T}(t-s)
\right)
\,\mathrm{d}s.
\label{eq:cov_solution}
\end{align}

An important direction for future work is to extend latent evolution beyond the OU family to richer families of FP equations, such as Gaussian mixtures or other finite-dimensional statistical manifolds, with the goal of enabling more expressive latent representations while preserving as much analytical tractability as possible. However, these extensions may not, in general, admit closed-form probability densities over time.

\subsection{Transport map under the Knothe--Rosenblatt rearrangement}
\label{subsec:transport_map}

The latent stochastic evolution determines the intrinsic stochastic dynamics of the system. The remaining degree of freedom lies in the transport map connecting the latent and observation spaces. Rather than parameterizing the transport map $T$, we parameterize its inverse
\begin{equation}
S(\cdot,t)=T^{-1}(\cdot,t),
\qquad
\forall\,t\in\mathcal T,
\label{eq:inverse_map}
\end{equation}
since the observation-space density is naturally evaluated through the inverse transformation via the change-of-variables formula.

To parameterize the inverse map, we adopt the Knothe--Rosenblatt (KR) rearrangement~\cite{Rosenblatt1952,Knothe1957,Baptista2023ATM,han2026inference}, which represents an invertible transport through a triangular sequence of one-dimensional monotone transformations,
\begin{equation}
S(\boldsymbol{x}, t)
=
\begin{bmatrix*}[l]
S_1(x_1,t)\\
S_2(x_1,x_2,t)\\
\vdots\\
S_d(x_1,\ldots,x_d,t)
\end{bmatrix*},
\label{eq:kr_map}
\end{equation}
The KR rearrangement represents the inverse transport as a triangular composition of one-dimensional transformations. Consequently, its Jacobian is lower triangular, and the determinant is given by
$
\det\nabla S(\boldsymbol{x},t)
=
\prod_{i=1}^{d}
\frac{\partial S_i}{\partial x_i}
.
$
This structure enables efficient evaluation of the Jacobian determinant required by the change-of-variables formula while reducing the construction of a multivariate invertible map to a sequence of one-dimensional monotone transformations.
Each component of the KR rearrangement is parameterized by a monotone neural network,
\begin{equation}
S_i(\boldsymbol{x}_{1:i},t; \boldsymbol{\theta}_{S})
=
f_i(\boldsymbol{x}_{1:i-1},0,t; \boldsymbol{\theta}_{S})
+
\int_{0}^{x_i}
g
\!\left(
\frac{\partial f_i}{\partial s}
(\boldsymbol{x}_{1:i-1},s,t; \boldsymbol{\theta}_{S})
\right)
\,\mathrm{d}s,
\label{eq:kr_parameterization}
\end{equation}
where
$
\boldsymbol{x}_{1:i}
=
(x_1,\ldots,x_i),
$
$f_i:\mathbb{R}^{i}\times\mathcal{T}\rightarrow\mathbb{R}$ is represented by a neural network with parameter
$
\boldsymbol{\theta}_{S}.
$,
and $g:\mathbb{R}\rightarrow\mathbb{R}_{>0}$ is a strictly positive activation function.
Since
$
\frac{\partial S_i}{\partial x_i}
=
g
\!\left(
\frac{\partial f_i}{\partial s}
\right)
>
0,
$
each component $S_i$ is strictly monotone increasing with respect to its final argument $x_i$. In combination with the lower triangular structure of $\nabla_{\boldsymbol{x}} S$, this makes the KR rearrangement globally invertible, guaranteeing the existence and uniqueness of $T = S^{-1}$: Given a latent coordinate
$
\boldsymbol{z}\in\mathcal{Z},
$
the corresponding observation is recovered by solving
$
\boldsymbol{z}=S(\boldsymbol{x},t).
$
Because the KR map is lower triangular and each component is strictly monotone with respect to its final argument, the inversion reduces to a sequence of one-dimensional nonlinear equations, each of which admits a unique solution.

The parameterized inverse map therefore determines both the transport map and its Jacobian determinant, providing all quantities required to evaluate observation-space likelihoods in the probabilistic inference framework developed in the following subsection.

\subsection{Joint estimation of latent dynamics and discrepancy transport map}
\label{subsec:joint_estimation}

The latent stochastic dynamics and the transport map together determine the probability distributions observed in the data. Under the Gaussian initial condition assumed in the previous subsection, the latent stochastic evolution is uniquely determined by the initial mean and covariance together with the OU parameters through the analytical solution in \cref{eq:mean_solution,eq:cov_solution}. In practice, the initial mean and covariance are estimated empirically from the observed samples at the initial observation time. Consequently, the unknowns of the inverse problem reduce to the latent dynamical parameters and the transport-map parameters
$
\boldsymbol{\theta}
=
\left(
\boldsymbol{\theta}_{\mathrm{OU}},
\boldsymbol{\theta}_{S}
\right)
$.
Given the latent stochastic evolution, the predicted observation-space probability density is obtained through the transport map,
$
\rho^{\mathrm{pred}}(\boldsymbol{x},t)
=
\left(
T_{\sharp}(\cdot,t)\zeta(\cdot,t)
\right)(\boldsymbol{x}).
$
Since the inverse transport map
$
S(\cdot,t)
=
T^{-1}(\cdot,t)
$
is parameterized explicitly, the predicted observation-space density is evaluated using the change-of-variables formula
\begin{equation}
\rho^{\mathrm{pred}}(\boldsymbol{x},t)
=
\left(
T_{\sharp}(\cdot,t)\zeta(\cdot,t)
\right)
(\boldsymbol{x})
=
\zeta
\!\left(
S(\boldsymbol{x},t; \boldsymbol{\theta}_{S}),
t;
\boldsymbol{\theta}_{\mathrm{OU}}
\right)
\left|
\det\nabla
S(\boldsymbol{x},t;\boldsymbol{\theta}_{S})
\right|.
\label{eq:pushforward_density}
\end{equation}

\paragraph{Distribution matching}

The primary objective is to match the predicted observation-space distributions to the observed snapshot distributions. We measure this discrepancy using the Kullback--Leibler (KL) divergence
\begin{equation}
\mathcal{L}_{\mathrm{KL}}
=
\sum_{k=0}^{n}
D_{\mathrm{KL}}
\!\left(
\rho_{t_k}
\,\middle\|\,
\rho_{t_k}^{\mathrm{pred}}
\right).
\label{eq:kl_objective}
\end{equation}
Using the change-of-variables formula \cref{eq:pushforward_density}, the KL divergence at $t_k$ becomes
\begin{equation}
D_{\mathrm{KL}}
\!\left(
\rho_{t_k}
\,\middle\|\,
\rho_{t_k}^{\mathrm{pred}}
\right)
=
\int_{\mathcal{X}}
\rho(\boldsymbol{x},t_k)
\log
\frac{
\rho(\boldsymbol{x},t_k)
}{
\zeta
(
S(\boldsymbol{x},t_k; \boldsymbol{\theta}_{S}),
t_k;
\boldsymbol{\theta}_{\mathrm{OU}}
)
\left|
\det\nabla
S(\boldsymbol{x},t_k;\boldsymbol{\theta}_{S})
\right|
}
\,\mathrm d\boldsymbol{x}.
\end{equation}
Since the entropy of
$\rho_{t_k}$
is independent of the optimization variables, minimizing
$\mathcal L_{\mathrm{KL}}$
is equivalent to minimizing
\begin{equation}
\mathcal L_{\mathrm{data}}
=
\sum_{k=0}^{n}
\mathbb E_{\rho_{t_k}}
\left[
-
\log
\zeta
(
S(\boldsymbol{x},t_k; \boldsymbol{\theta}_{S}),
t_k;
\boldsymbol{\theta}_{\mathrm{OU}}
)
-
\log
\left|
\det\nabla
S(\boldsymbol{x},t_k; \boldsymbol{\theta}_{S})
\right|
\right].
\label{eq:data_objective}
\end{equation}

\paragraph{Latent regularization}
The OU process provides an analytically tractable but intentionally restrictive latent stochastic model. Since the latent and observation spaces are identified with the same ambient coordinate system, we may evaluate the latent density directly on the observed samples before introducing geometric deformation through the transport map. To encourage the latent dynamics to explain as much of the observed probability evolution as possible, we additionally minimize the latent fitting objective
\begin{equation}
\mathcal L_{\mathrm{latent}}
=
-
\sum_{k=0}^{n}
\mathbb E_{\rho_{t_k}}
\left[
\log
\zeta
\!\left(
\boldsymbol{x},
t_k;
\boldsymbol{\theta}_{\mathrm{OU}}
\right)
\right].
\label{eq:latent_loss}
\end{equation}
This term encourages the latent stochastic dynamics to capture the intrinsic probability evolution in OU form, leaving only the residual geometric discrepancy to be represented by the discrepancy transport map.

\paragraph{Elastic regularization}

As established in the previous subsection, the latent--transport decomposition is generally non-unique. To enforce uniqueness of the decomposition, we regularize the transport map using an elastic deformation energy
\begin{equation}
\mathcal L_{\mathrm{el}}
=
\sum_{k=0}^{n}
E_{\mathrm{el}}
\!\left(
S^{-1}(\cdot,t_k; \boldsymbol{\theta}_{S})
\right),
\label{eq:elastic_regularization}
\end{equation}
where
$E_{\mathrm{el}}$
denotes the elastic energy function. A range of forms is available for the elastic energy, and they have been widely studied in the nonlinear solid mechanics literature \cite{ogden1997non,holzapfel2002nonlinear,bigoni2012nonlinear}. If the transport of interest is of a tracer mass through a solid continuum, a form suitable to that material may be chosen.
In the present work, we adopt the St.\ Venant--Kirchhoff (SVK) hyperelastic model.

Let
\begin{equation}
\boldsymbol{F}
=
\nabla_{\boldsymbol{z}}T(\boldsymbol{z},t)
=
\left(
\nabla_{\boldsymbol{x}}S(\boldsymbol{x},t)
\right)^{-1}
\label{eq:deformation_gradient}
\end{equation}
denote the deformation gradient associated with the transport map. The corresponding right Cauchy--Green tensor and Green--Lagrange strain tensor are
\begin{equation}
\boldsymbol{C}
=
\boldsymbol{F}^{\mathsf T}\boldsymbol{F},
\qquad
\boldsymbol{E}
=
\frac12
\left(
\boldsymbol{C}
-
\boldsymbol{I}
\right),
\label{eq:green_lagrange}
\end{equation}
where
$\boldsymbol{I}$
is the identity tensor. The SVK strain-energy density is given by
\begin{equation}
W(\boldsymbol{F})
=
\frac12
\left[
\lambda
\left(
\operatorname{tr}\boldsymbol{E}
\right)^2
+
\mu\,
\boldsymbol{E}:\boldsymbol{E}
\right],
\label{eq:stvk_energy}
\end{equation}
where
$\lambda$
and
$\mu$
are the Lam\'e parameters.

\paragraph{Overall optimization}

The three objectives are combined into the optimization problem
\begin{equation}
\min_{\boldsymbol{\theta}}
\;
\mathcal L_{\mathrm{data}}
+
\beta
\mathcal L_{\mathrm{latent}}
+
\gamma
\mathcal L_{\mathrm{el}},
\label{eq:joint_optimization}
\end{equation}
where
$\beta$
and
$\gamma$
control the relative contributions of latent model fitting and elastic regularization, respectively.

Since the observed probability distributions are available only through samples,
$
\left\{
\boldsymbol{x}_{t_k}^{(i)}
\right\}_{i=1}^{m_k},
\allowbreak
k=0,\ldots,n,
$
the expectation operators in
\cref{eq:latent_loss,eq:data_objective}
are approximated by empirical averages over the observed samples.

\section{Numerical examples}
\label{sec:numerical_results}

This section evaluates the proposed framework on two synthetic examples designed to validate complementary aspects of the latent--transport decomposition. In both examples, the ground-truth stochastic dynamics are simulated using the Euler--Maruyama scheme~\cite{KloedenPlaten1992}, while the observed data consist only of independent snapshot samples collected at discrete observation times, with no correspondence retained between samples across time. The latent stochastic evolution is modeled analytically by the OU process introduced in \cref{subsec:latent_dynamics}, whereas the discrepancy transport map is parameterized using the KR rearrangement described in \cref{subsec:transport_map}. The latent dynamical parameters and transport-map parameters are jointly estimated according to the optimization problem in \cref{subsec:joint_estimation}.

\subsection{Annular quartic potential}
\label{subsec:quartic_potential}

The first example considers the proposed framework on an annular quartic potential. The objective is to determine whether the framework can recover a simple latent stochastic evolution from snapshot observations when the observed probability distribution exhibits a nontrivial annular geometry but does not require substantial geometric correction. In this setting, the discrepancy transport map is expected to provide only a moderate deformation, allowing the latent OU dynamics to explain most of the probability evolution.

\paragraph{Problem setup}
We consider an annular quartic potential on $\mathbb{R}^2$. The ground-truth observation-space potential $\psi_{\boldsymbol{x}}$, diffusion tensor $\boldsymbol{D}_{\boldsymbol{x}}$, and initial distribution $\rho_0$ are given by
\begin{align}
\psi_{\boldsymbol{x}}(\boldsymbol{x})
&=
\frac{1}{4}
\left(
\boldsymbol{x}^{\mathsf T}\boldsymbol{x}
\right)^2
-
\frac{1}{2}
\boldsymbol{x}^{\mathsf T}
\begin{bmatrix}
1.5 & 0 \\
0 & 1.5
\end{bmatrix}
\boldsymbol{x},
\label{eq:annular_potential}
\\
\boldsymbol{D}_{\boldsymbol{x}}
&=
\begin{bmatrix}
0.1 & 0 \\
0 & 0.2
\end{bmatrix},
\label{eq:annular_diffusion}
\\
\rho_0
&=
\mathcal{N}\!\left(
\begin{bmatrix}
0 \\
0
\end{bmatrix},
\begin{bmatrix}
0.1 & 0 \\
0 & 0.1
\end{bmatrix}
\right).
\label{eq:annular_initial}
\end{align}
The process is simulated over the time interval $[0,2]$ using the Euler--Maruyama scheme. At each observation time, only independent snapshot samples drawn from the evolving distribution are retained; although the simulator generates trajectories, their correspondence across observation times is not preserved in the observed data.

\paragraph{Results}

\begin{figure}[htbp]
    \centering
    \begin{subfigure}{\textwidth}
        \centering
        \makebox[0.19\textwidth][c]{\includegraphics[height=0.11\textheight]{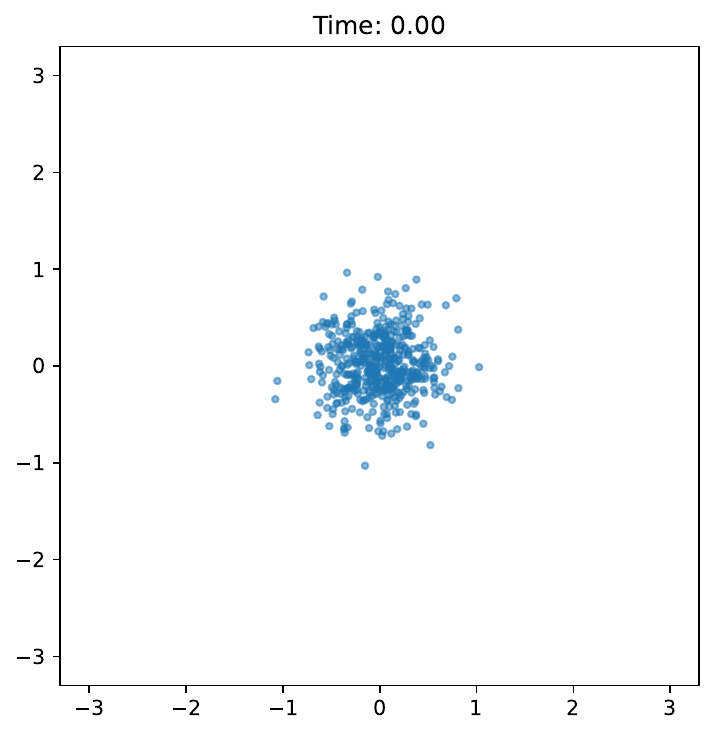}}
        \makebox[0.19\textwidth][c]{\includegraphics[height=0.11\textheight]{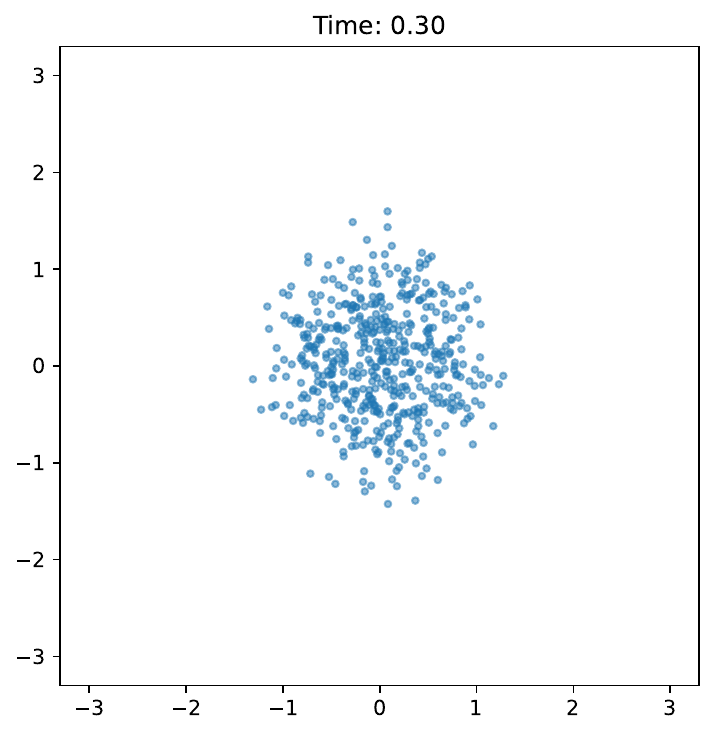}}
        \makebox[0.19\textwidth][c]{\includegraphics[height=0.11\textheight]{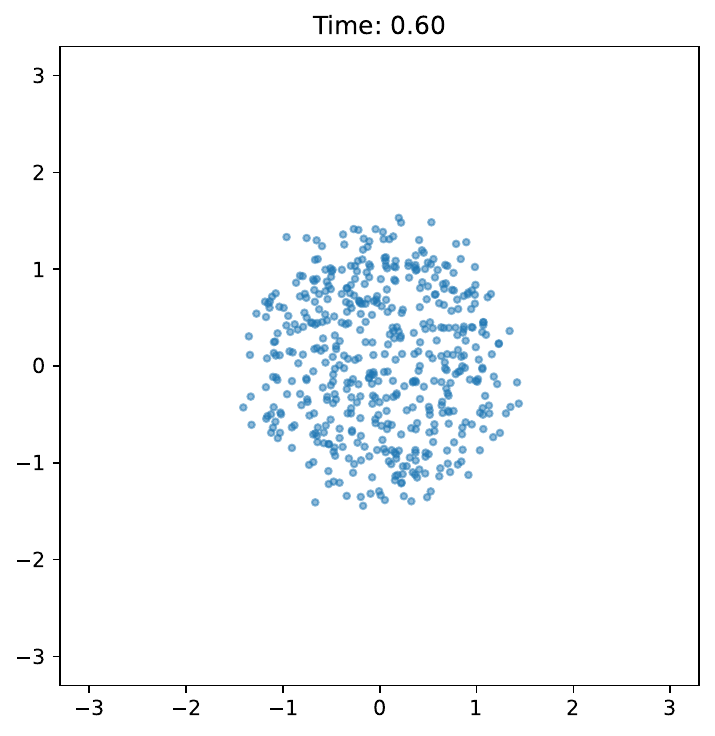}}
        \makebox[0.19\textwidth][c]{\includegraphics[height=0.11\textheight]{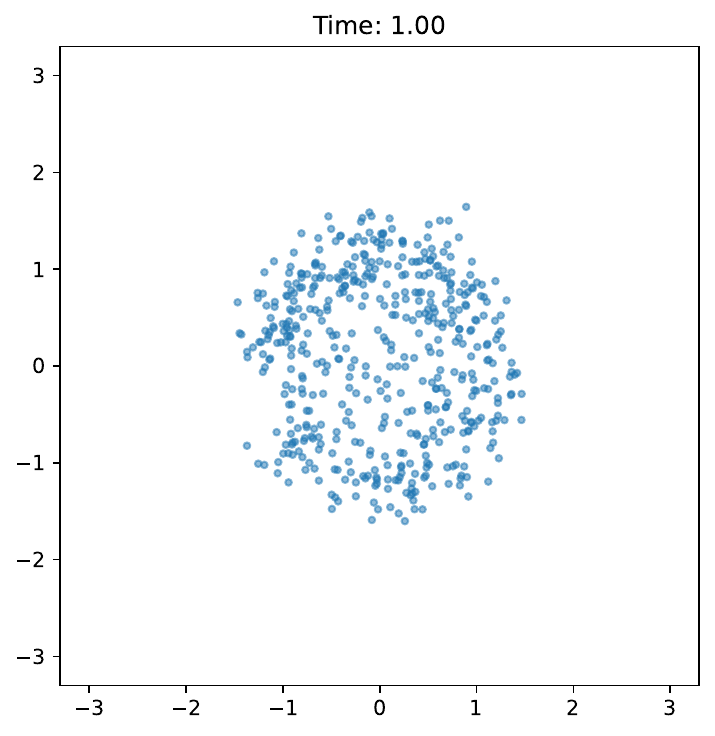}}
        \makebox[0.19\textwidth][c]{\includegraphics[height=0.11\textheight]{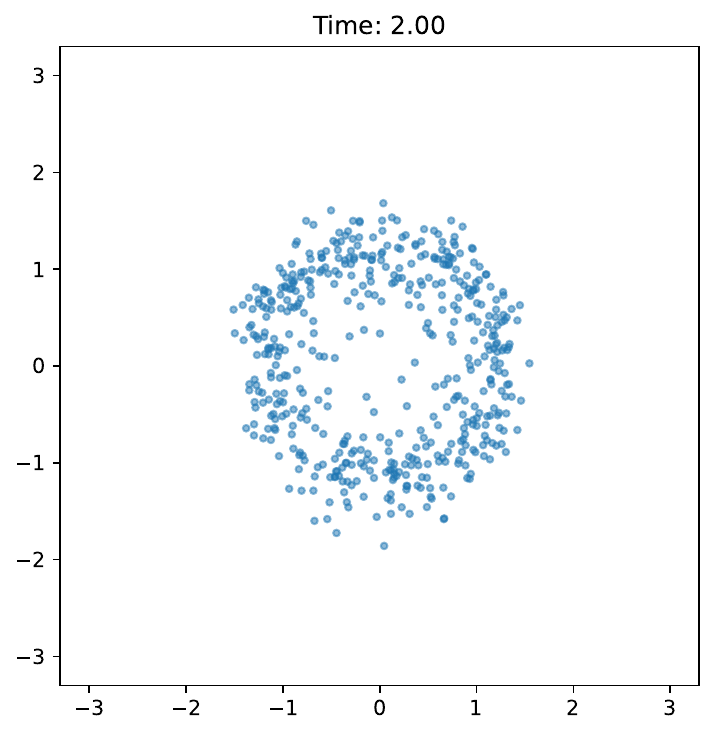}}
        \caption{Observed snapshot distributions.}
    \end{subfigure}
    \begin{subfigure}{\textwidth}
        \centering
        \makebox[0.19\textwidth][c]{\includegraphics[height=0.11\textheight]{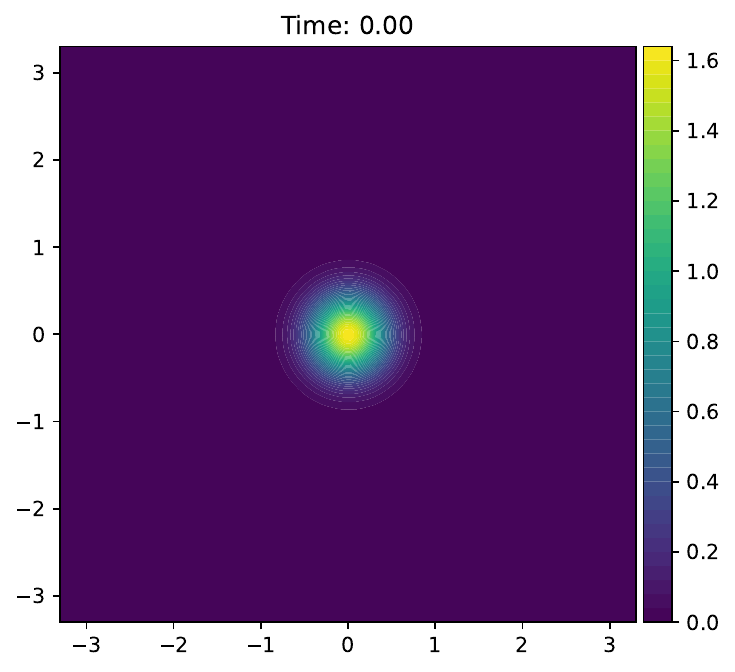}}
        \makebox[0.19\textwidth][c]{\includegraphics[height=0.11\textheight]{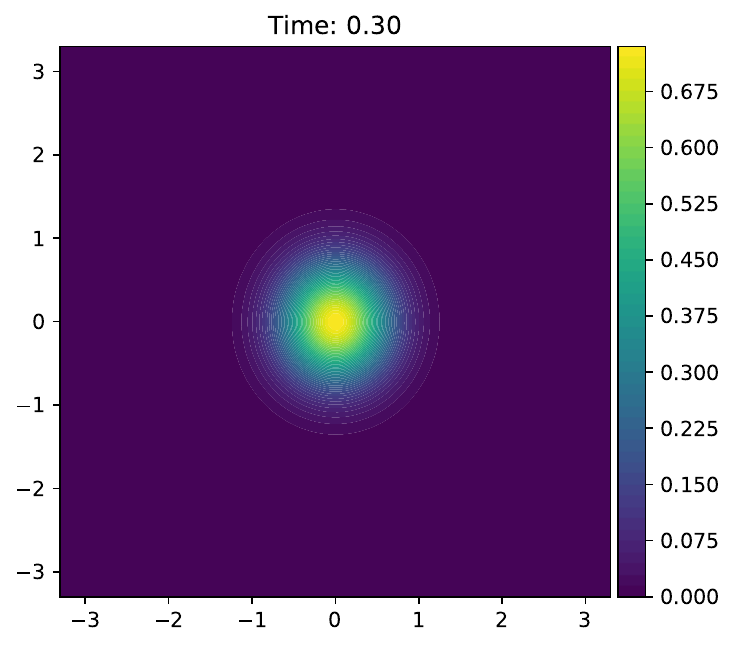}}
        \makebox[0.19\textwidth][c]{\includegraphics[height=0.11\textheight]{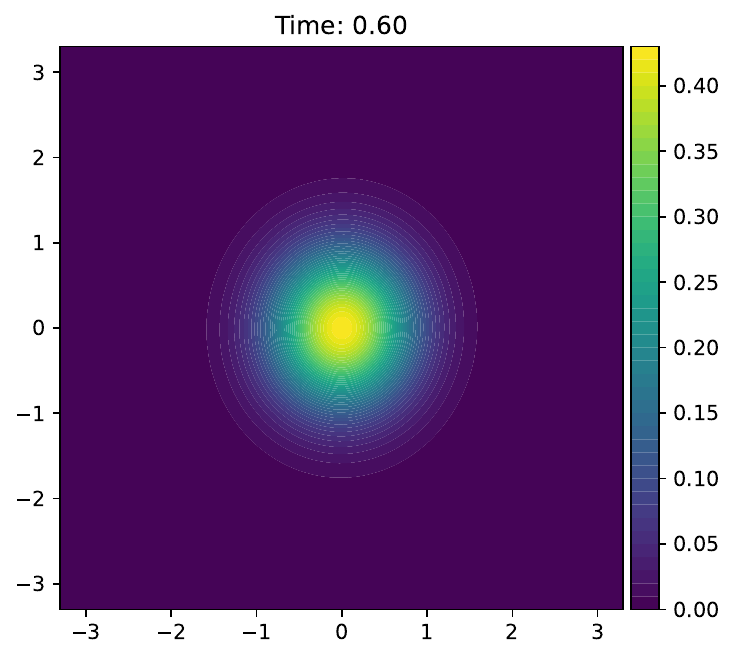}}
        \makebox[0.19\textwidth][c]{\includegraphics[height=0.11\textheight]{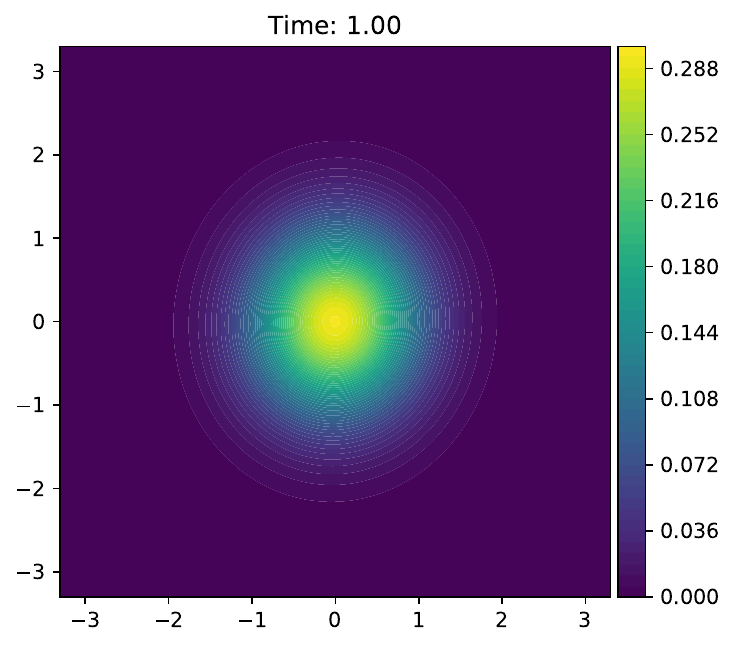}}
        \makebox[0.19\textwidth][c]{\includegraphics[height=0.11\textheight]{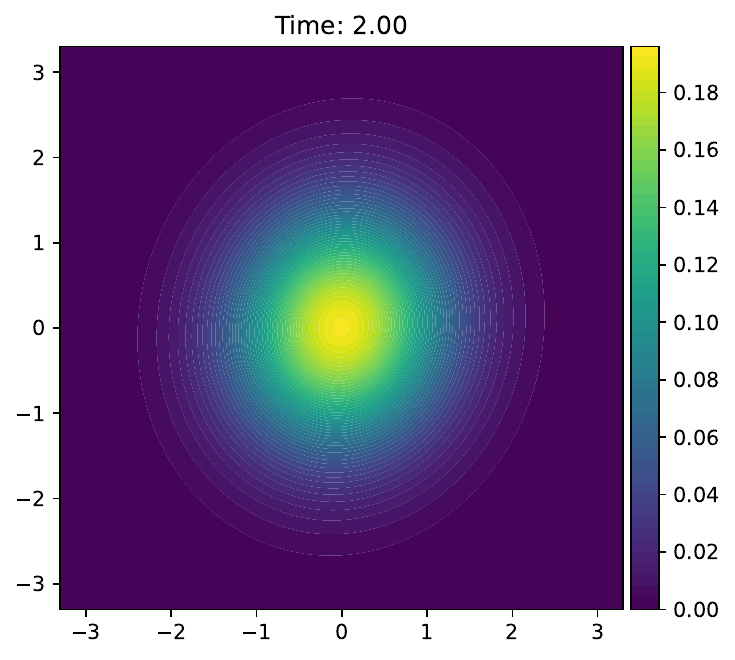}}
        \caption{Latent probability distributions predicted by the OU dynamics.}
    \end{subfigure}
    \begin{subfigure}{\textwidth}
        \centering
        \makebox[0.19\textwidth][c]{\includegraphics[height=0.11\textheight]{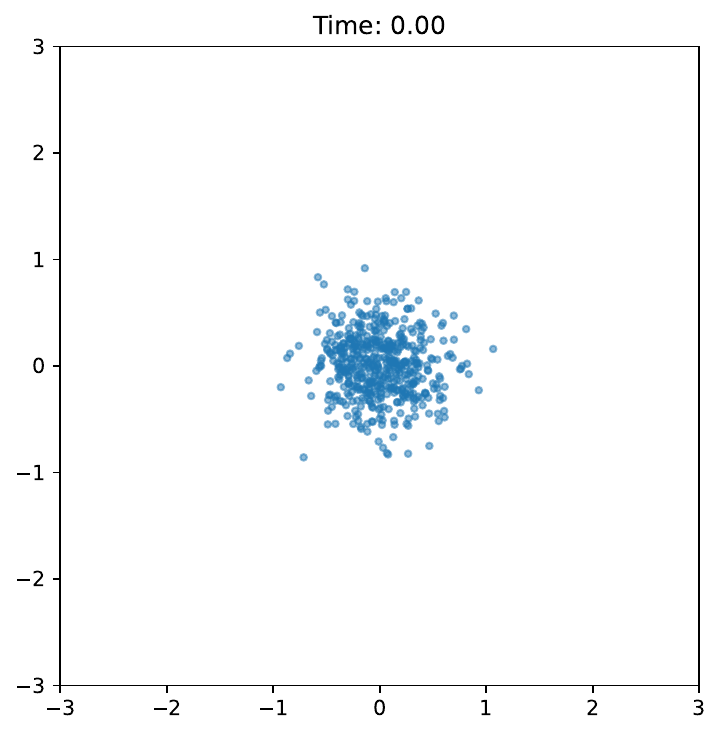}}
        \makebox[0.19\textwidth][c]{\includegraphics[height=0.11\textheight]{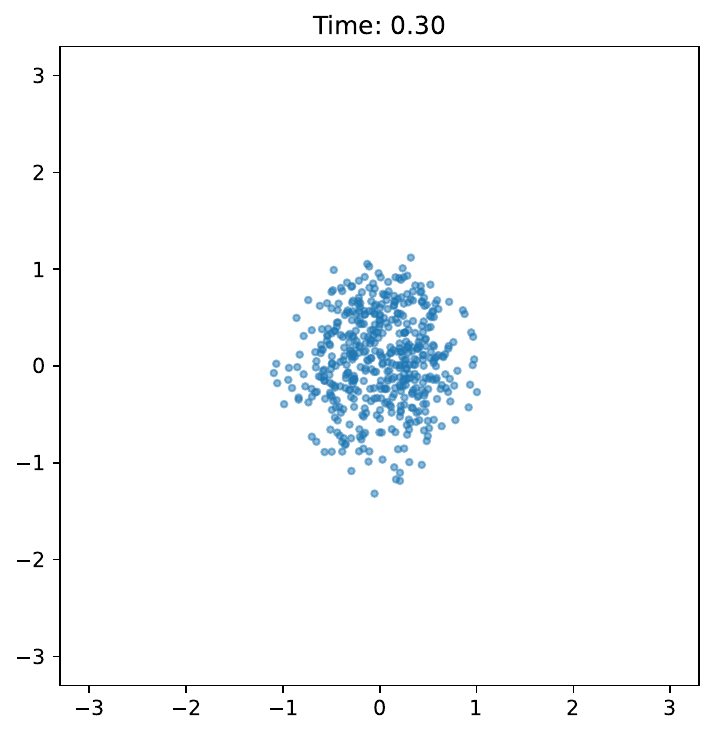}}
        \makebox[0.19\textwidth][c]{\includegraphics[height=0.11\textheight]{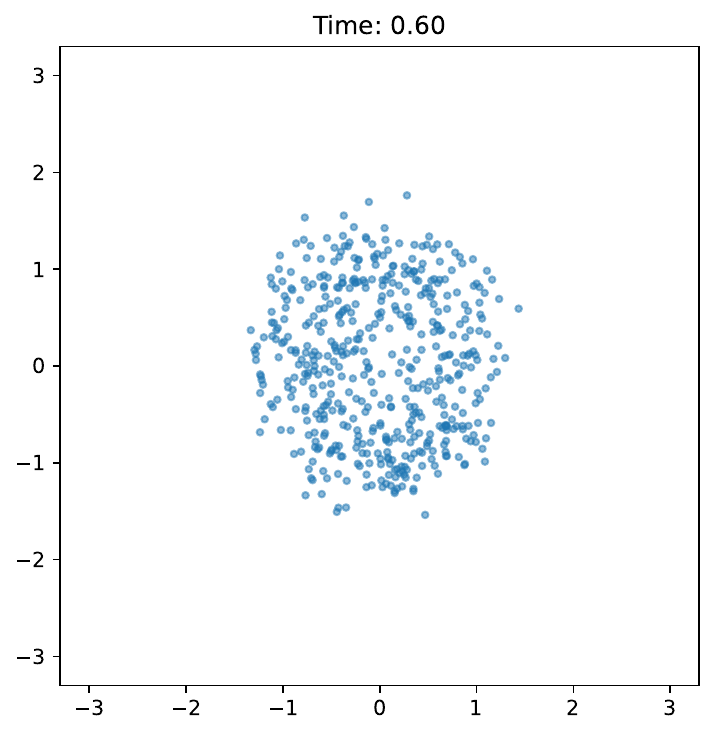}}
        \makebox[0.19\textwidth][c]{\includegraphics[height=0.11\textheight]{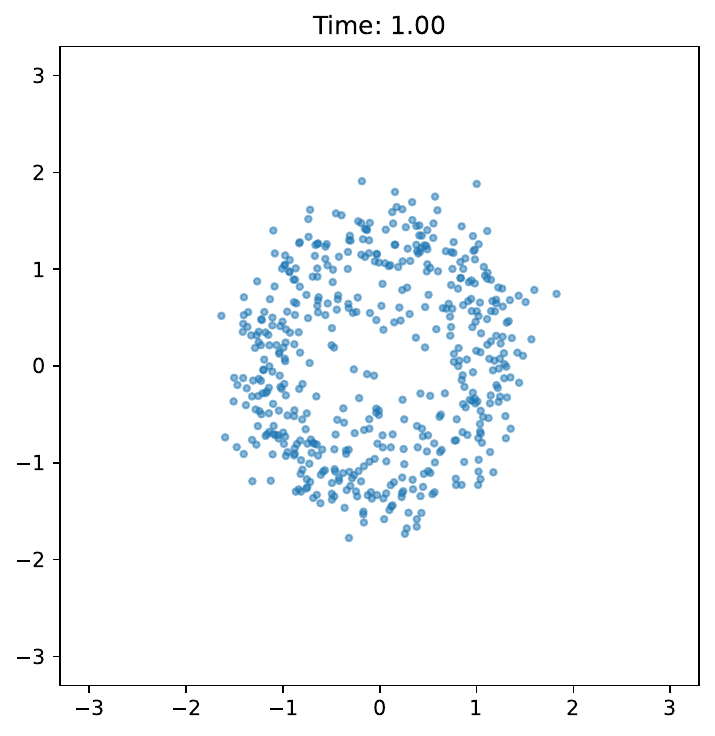}}
        \makebox[0.19\textwidth][c]{\includegraphics[height=0.11\textheight]{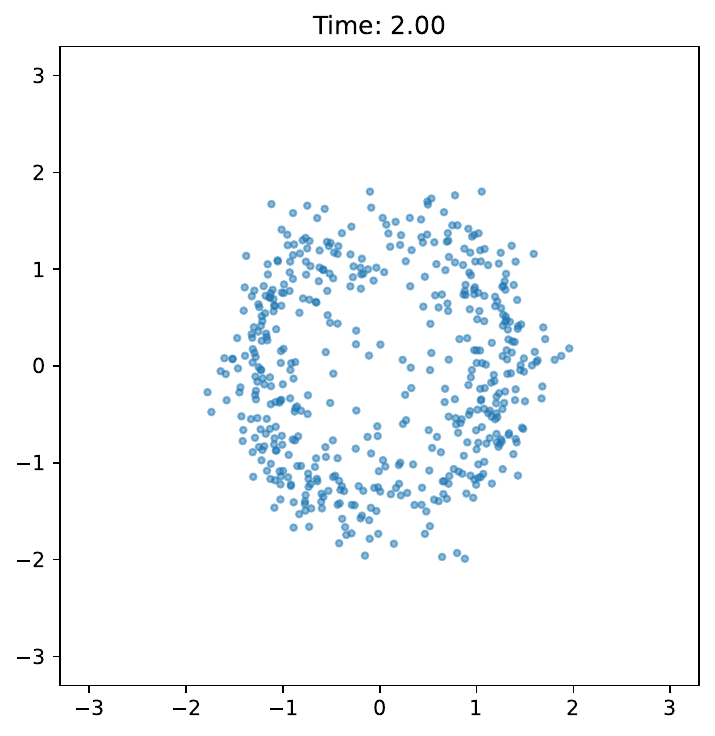}}
        \caption{Reconstructed observation-space distributions obtained by the learned transport map.}
    \end{subfigure}
    \caption{Annular quartic potential. Representative snapshots of the observed distributions, the inferred latent distributions, and the reconstructed observation-space distributions over the interval $[0,2]$.}
    \label{fig:quartic_results}
\end{figure}

\Cref{fig:quartic_results} summarizes the inferred latent--transport decomposition. The top row shows the observed snapshot distributions, the middle row shows the inferred latent probability distributions predicted analytically by the learned OU dynamics, and the bottom row shows the reconstructed observation-space distributions obtained by pushing the latent distributions through the learned discrepancy transport map.

The close agreement between the observed and reconstructed distributions demonstrates that the proposed framework reproduces the observed probability evolution while maintaining a simple latent stochastic model. In this example, the inferred latent dynamics account for most of the temporal evolution, whereas the discrepancy transport map provides only a moderate radial deformation that transforms the latent stochastic evolution into the observed annular probability distributions.

\subsection{Bimodal Rosenbrock potential}
\label{subsec:rosenbrock_potential}

The second example considers a substantially more challenging bimodal Rosenbrock potential. Unlike the previous example, the observed probability distributions exhibit both multimodality and strong nonlinear geometric deformation. The objective is to determine whether the proposed framework can preserve a simple latent stochastic evolution while allowing the discrepancy transport map to capture the complex geometry of the observed probability distributions.

\paragraph{Problem setup}

We consider a bimodal Rosenbrock potential on $\mathbb{R}^2$. The ground-truth potential is a soft minimum of two rotated Rosenbrock-type wells, given by
\begin{align}
\psi(\boldsymbol{x})
&=
-\tau\log\!\left(
e^{-\psi_1/\tau}
+
e^{-\psi_2/\tau}
\right),
\label{eq:rosenbrock_potential}
\\
\psi_i
&=
\alpha\left(\frac{u_i}{s}\right)^2
+
\beta\left(
v_i-\left(\frac{u_i}{s}\right)^2
\right)^2,
\qquad i=1,2,
\label{eq:rosenbrock_components}
\\
\begin{bmatrix}
u_i \\
v_i
\end{bmatrix}
&=
R(\theta_i)^{\mathsf T}
\left(
\boldsymbol{x}-\boldsymbol{c}_i
\right),
\qquad
R(\theta_i)
=
\begin{bmatrix}
\cos\theta_i & -\sin\theta_i \\
\sin\theta_i & \cos\theta_i
\end{bmatrix},
\qquad i=1,2.
\label{eq:rosenbrock_rotation}
\end{align}
The parameter values are
\begin{align}
\alpha &= 0.15, &
\beta &= 4.0, &
s &= 1.4, &
\tau &= 0.2, \nonumber\\
\boldsymbol{c}_1 &= (-0.6,-0.7)^{\mathsf T}, &
\boldsymbol{c}_2 &= (0.6,0.7)^{\mathsf T}, &
\theta_1 &= 0, &
\theta_2 &= \pi.
\label{eq:rosenbrock_parameters}
\end{align}
The diffusion tensor and initial distribution are
\begin{align}
\boldsymbol{D}_{\boldsymbol{x}}
&=
\begin{bmatrix}
0.1 & 0 \\
0 & 0.2
\end{bmatrix},
\label{eq:rosenbrock_diffusion}
\\
\rho_0
&=
\mathcal{N}\!\left(
\begin{bmatrix}
0 \\
0
\end{bmatrix},
\begin{bmatrix}
0.1 & 0 \\
0 & 0.1
\end{bmatrix}
\right).
\label{eq:rosenbrock_initial}
\end{align}
The process is simulated over the interval $[0,0.25]$ using the Euler--Maruyama scheme. As in the first example, only independent snapshot samples are retained at each observation time, and the trajectory correspondence generated by the simulator is discarded in the observed data.

\paragraph{Results}

\begin{figure}[htbp]
    \centering
    \begin{subfigure}{\textwidth}
        \centering
        \makebox[0.19\textwidth][c]{\includegraphics[height=0.11\textheight]{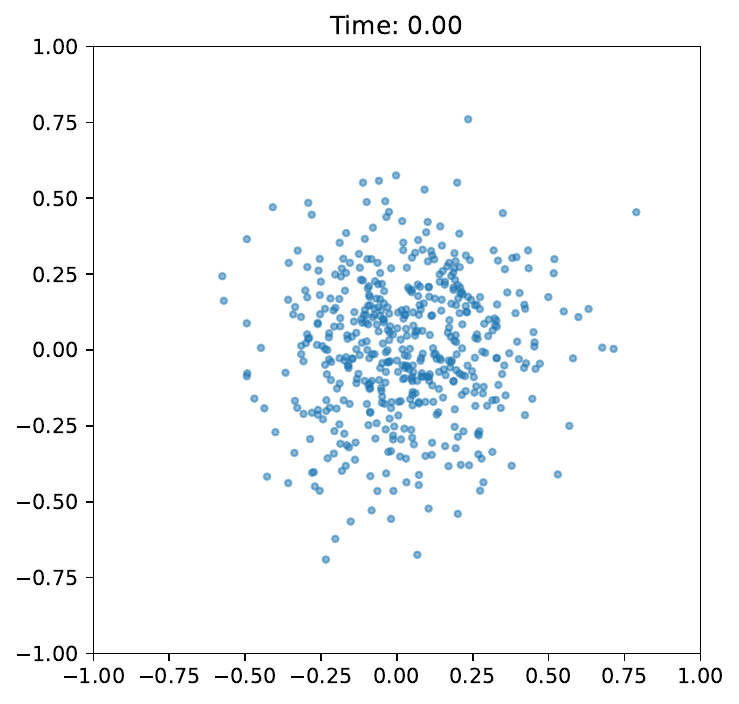}}
        \makebox[0.19\textwidth][c]{\includegraphics[height=0.11\textheight]{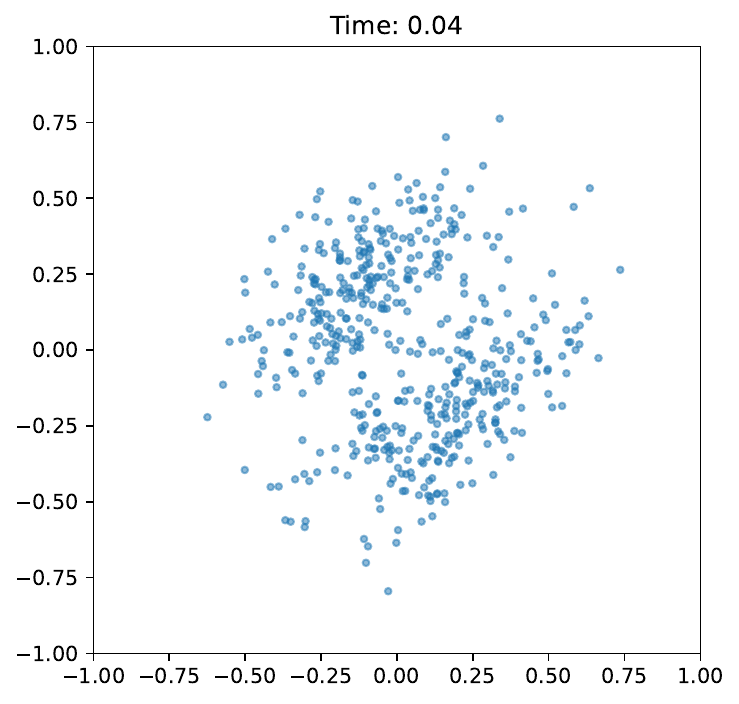}}
        \makebox[0.19\textwidth][c]{\includegraphics[height=0.11\textheight]{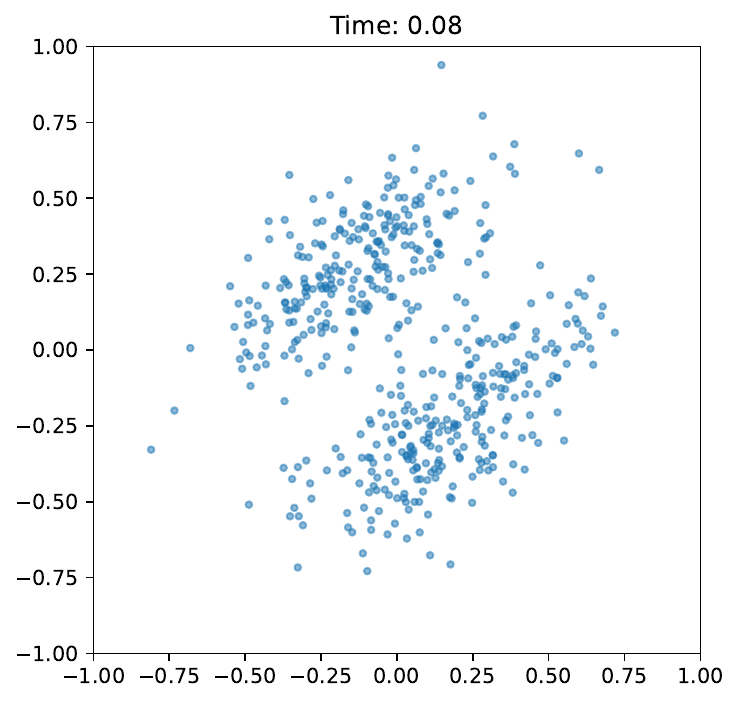}}
        \makebox[0.19\textwidth][c]{\includegraphics[height=0.11\textheight]{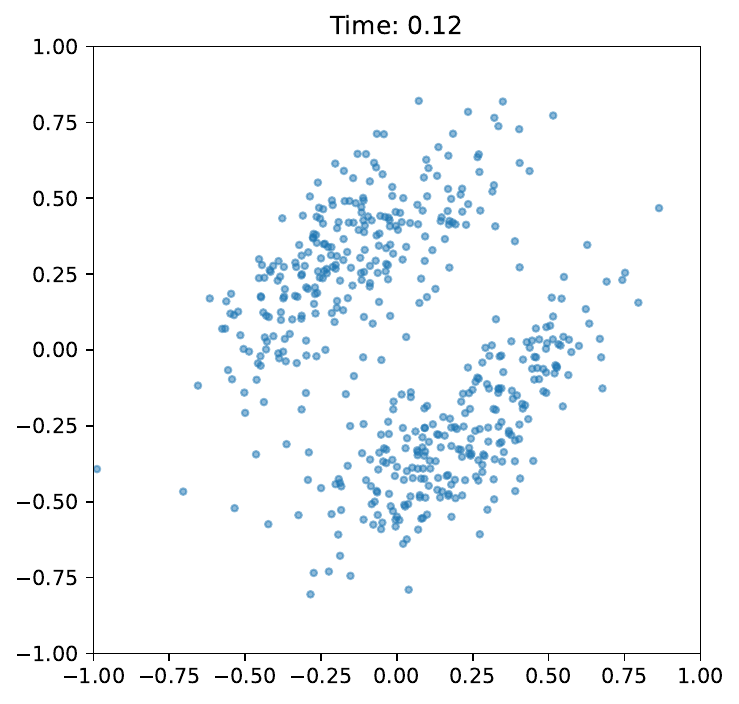}}
        \makebox[0.19\textwidth][c]{\includegraphics[height=0.11\textheight]{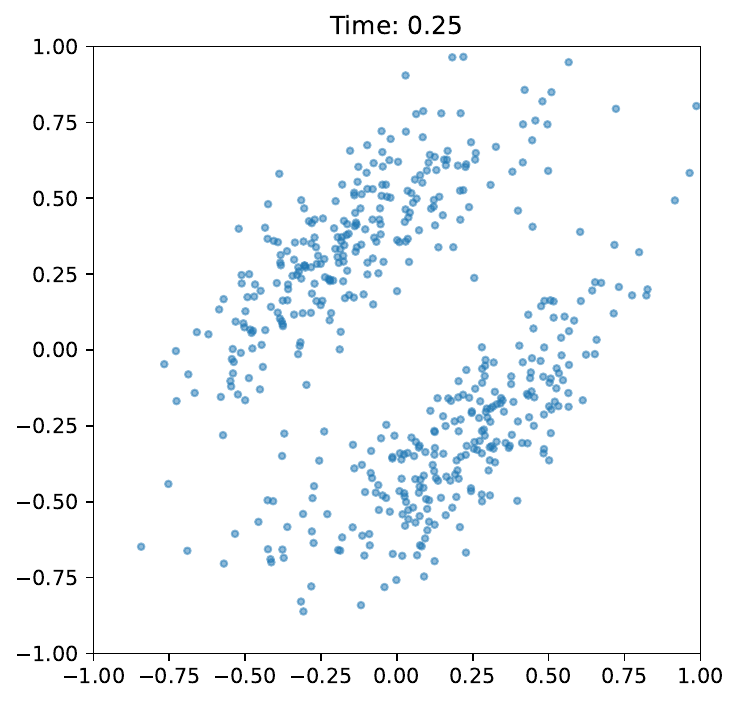}}
        \caption{Observed snapshot distributions.}
    \end{subfigure}
    \begin{subfigure}{\textwidth}
        \centering
        \makebox[0.19\textwidth][c]{\includegraphics[height=0.11\textheight]{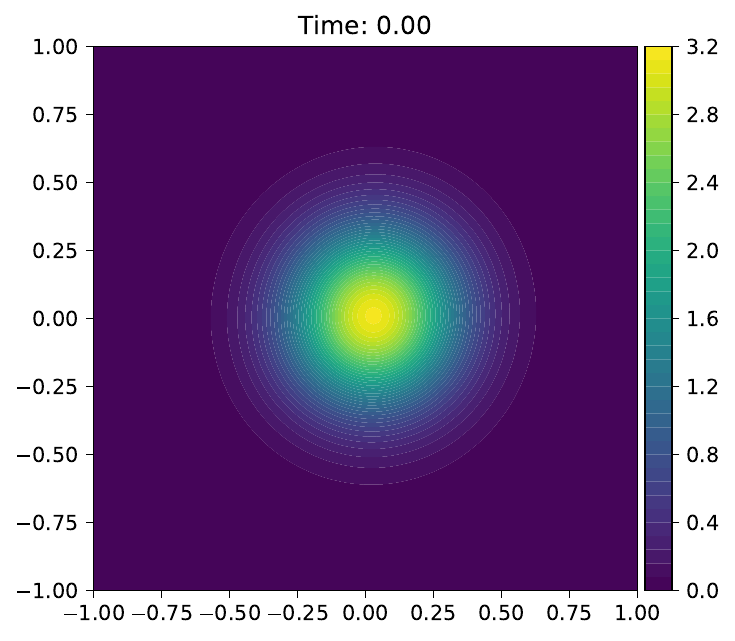}}
        \makebox[0.19\textwidth][c]{\includegraphics[height=0.11\textheight]{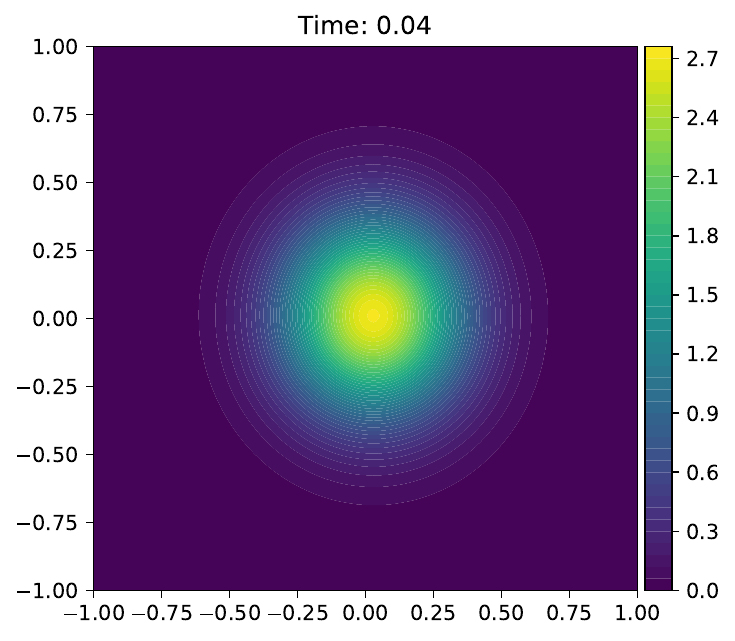}}
        \makebox[0.19\textwidth][c]{\includegraphics[height=0.11\textheight]{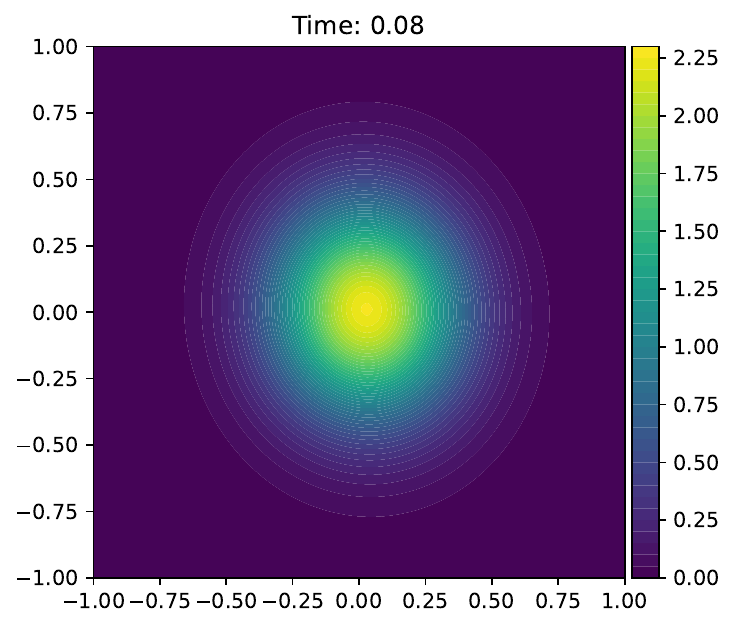}}
        \makebox[0.19\textwidth][c]{\includegraphics[height=0.11\textheight]{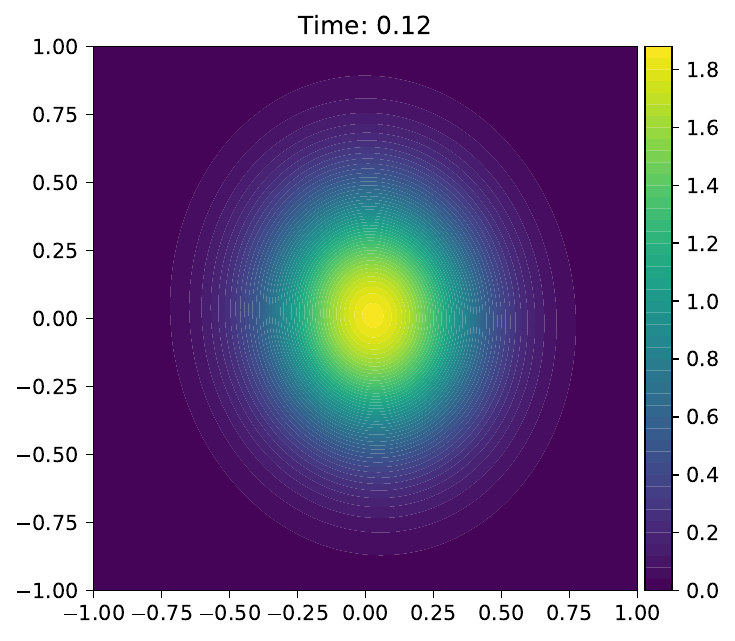}}
        \makebox[0.19\textwidth][c]{\includegraphics[height=0.11\textheight]{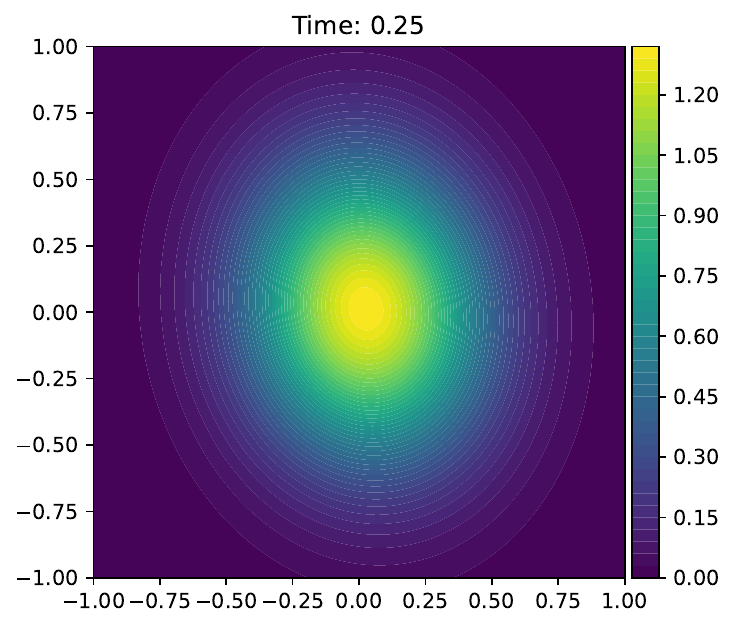}}
        \caption{Latent probability distributions predicted by the OU dynamics.}
    \end{subfigure}
    \begin{subfigure}{\textwidth}
        \centering
        \makebox[0.19\textwidth][c]{\includegraphics[height=0.11\textheight]{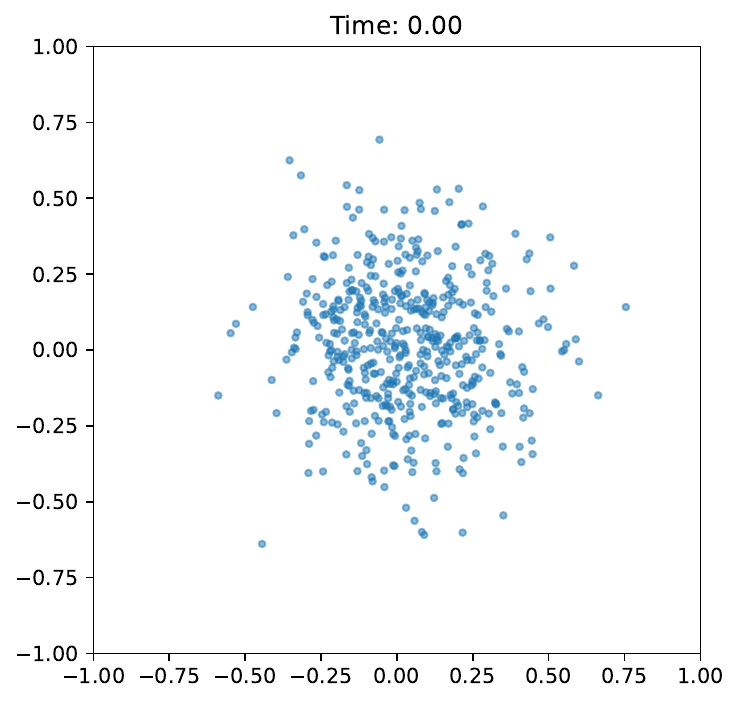}}
        \makebox[0.19\textwidth][c]{\includegraphics[height=0.11\textheight]{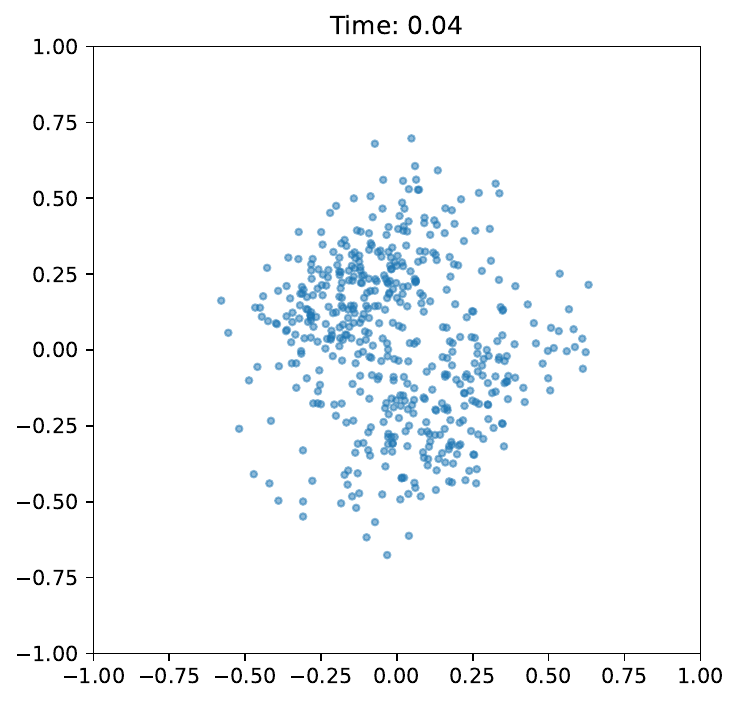}}
        \makebox[0.19\textwidth][c]{\includegraphics[height=0.11\textheight]{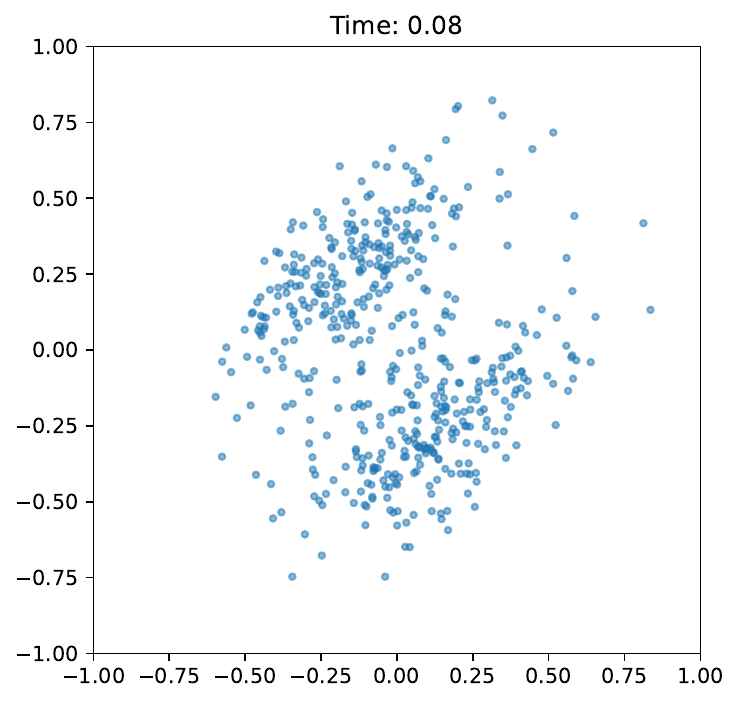}}
        \makebox[0.19\textwidth][c]{\includegraphics[height=0.11\textheight]{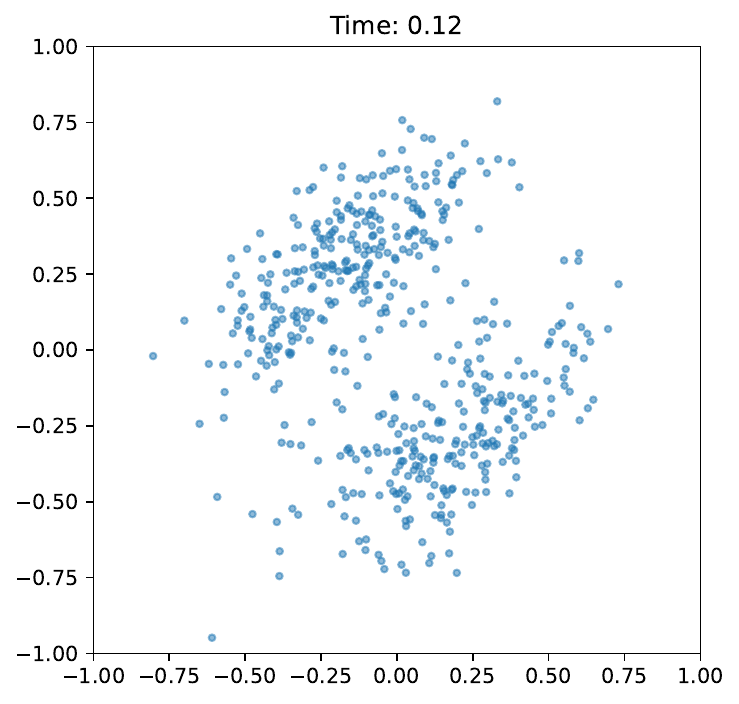}}
        \makebox[0.19\textwidth][c]{\includegraphics[height=0.11\textheight]{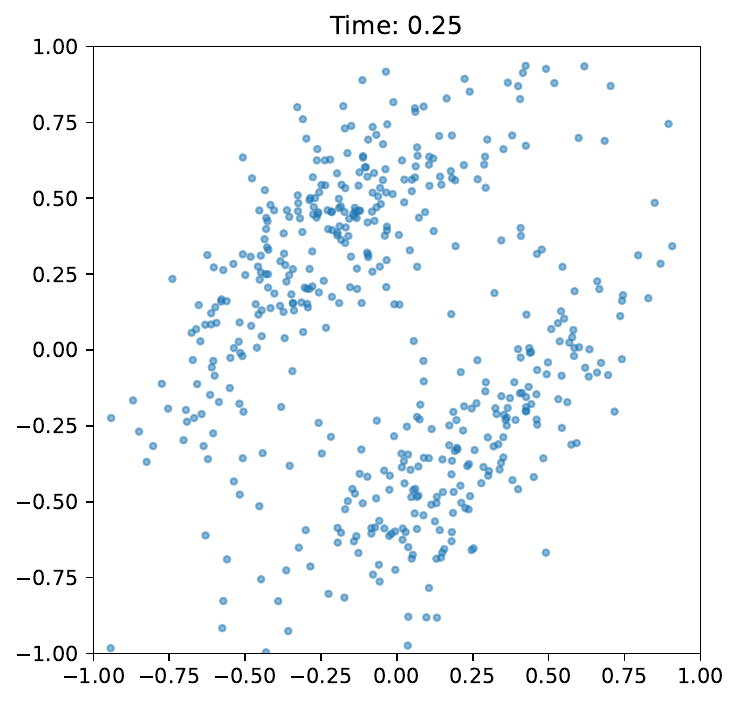}}
        \caption{Reconstructed observation-space distributions obtained by the learned transport map.}
    \end{subfigure}
    \caption{Bimodal Rosenbrock potential example. Representative snapshots of the observed distributions, the inferred latent distributions, and the reconstructed observation-space distributions over the interval $[0,0.25]$.}
    \label{fig:rosenbrock_results}
\end{figure}

\Cref{fig:rosenbrock_results} presents the inferred latent and reconstructed probability distributions. Despite the substantial geometric complexity of the observed distributions, the inferred latent probability distributions remain simple under the OU dynamics, while the discrepancy transport map captures the bimodal structure together with the strongly curved probability contours.

This example illustrates the central idea of the proposed framework. The latent stochastic dynamics model the intrinsic temporal evolution in a canonical FP form, whereas the discrepancy transport map accounts for the nonlinear geometric discrepancy between the latent and observation spaces. The learned transport transforms the simple latent stochastic evolution into the observed bimodal Rosenbrock distributions without requiring an equally complex latent stochastic model.

\subsection{Summary of numerical results}

Taken together, these two examples demonstrate the complementary roles of the latent stochastic dynamics and the discrepancy transport map. The annular quartic potential shows that when the observed probability evolution is geometrically simple, the OU model explains most of the dynamics and only modest transport correction is required. The bimodal Rosenbrock potential demonstrates that substantially more complex observed probability distributions can still be represented using the same simple latent stochastic model, with the discrepancy transport map accounting for the nonlinear geometric structure.

Overall, the numerical experiments support the central hypothesis of this work: observed probability evolution can be interpreted as the combined effect of an intrinsic latent stochastic evolution governed by a canonical FP equation and a discrepancy transport map that captures the remaining geometric deformation.

\section{Discussion}
\label{sec:discussion}

\subsection{Interpretation of the latent--transport decomposition}

The proposed framework formulates population-level dynamical inference as an inverse problem over evolving probability measures. Rather than inferring from the trajectories of individual particles, the framework seeks a latent probability measure $\zeta_t$ and a discrepancy transport map $T(\cdot,t)$ such that
$
\rho_t = T_{\sharp}(\cdot,t)\zeta_t
$
from distribution snapshots.
In this decomposition, the latent measure represents the intrinsic stochastic evolution of the underlying system, while the transport map captures the geometric mismatch between the latent and observation spaces.

This decomposition is inherently non-unique: for a given sequence of observed probability distributions, many combinations of latent stochastic evolutions and transport maps may reproduce the observations. The purpose of the framework is therefore not to identify an arbitrary admissible decomposition, but to select the simplest physically meaningful one. In the present formulation, this selection is achieved through three modeling choices: an OU-constrained FP latent model, a KR transport parameterization, and an elastic regularization of the transport energy. Together, these assumptions bias the solution toward latent dynamics that explain as much of the temporal evolution as possible, while reserving the transport map for the residual geometric discrepancy.

\subsection{Connection to transport in elastic solids}

Suppose that $\mathcal{X},\mathcal{Z} \subset \mathbb{R}^3$ and $\zeta(\boldsymbol{z},t)$ is the density of a tracer mass undergoing transport in $\mathcal{Z}$ governed by the following canonical form of the FP equation:
\begin{equation}
\frac{\partial \zeta(\boldsymbol{z},t)}{\partial t}
=
\nabla\cdot
\big(
\underbrace{\nabla\psi(\boldsymbol{z})}_{-\boldsymbol{v}(\boldsymbol{z})}\zeta(\boldsymbol{z},t)
\big)
+
\frac12
\nabla\cdot
\left(
\boldsymbol{D}_{\boldsymbol{z}}\nabla\zeta(\boldsymbol{z},t)
\right).
\label{eq:fp_w_pot}
\end{equation}

In the observation space $\mathcal{X}$, the observed density of the tracer mass is $\rho(\boldsymbol{x},t)$. The use of the change-of-variables formula $\zeta(\boldsymbol{z},t)=\rho(T(\boldsymbol{z},t),t)\vert\det\nabla T(\boldsymbol{z},t)\vert$, the Piola transform
\begin{equation}
    \boldsymbol{n}_{\boldsymbol{x}} \mathrm{d}\Gamma_{\boldsymbol{x}} = \vert \det\nabla T\vert \,T^{-\top}\boldsymbol{n}_{\boldsymbol{z}} \mathrm{d}\Gamma_{\boldsymbol{z}}
    \label{eq:piola}
\end{equation}
to relate elemental area vectors $\boldsymbol{n}_{\boldsymbol{z}}\mathrm{d}\Gamma_{\boldsymbol{z}}$ and $\boldsymbol{n}_{\boldsymbol{x}}\mathrm{d}\Gamma_{\boldsymbol{x}}$ on surfaces $\Gamma_{\boldsymbol{z}},\Gamma_{\boldsymbol{x}} \subset\mathbb{R}^3$, and some tensor calculus leads to the transformed observation-space drift and diffusion,
\begin{align}
\boldsymbol{v}_{\boldsymbol{x}}(\boldsymbol{x},t)
&=
\frac{\partial T(\boldsymbol{z},t)}{\partial t} -
\nabla T(\boldsymbol{z},t)\nabla \psi(\boldsymbol{z})
\nonumber\\
&\qquad -
\frac{1}{2}
\nabla T(\boldsymbol{z},t)
\boldsymbol{D}_{\boldsymbol{z}}
\left(
\nabla T(\boldsymbol{z},t)^{-\top}
:
\nabla\nabla T(\boldsymbol{z},t)
\right),
\label{eq:transformed_drift}
\\
\boldsymbol{D}_{\boldsymbol{x}}(\boldsymbol{x},t)
&=
\nabla T(\boldsymbol{z},t)
\boldsymbol{D}_{\boldsymbol{z}}
\nabla T(\boldsymbol{z},t)^{-\top},
\label{eq:transformed_diffusion}
\end{align}
where $\boldsymbol{z}=T^{-1}(\boldsymbol{x},t)$ is implied on the right-hand sides. The observed density then satisfies
\begin{equation}
\frac{\partial\rho(\boldsymbol{x},t)}{\partial t}
 = -
\nabla_{\boldsymbol{x}}\!\cdot
\left(
\rho(\boldsymbol{x},t)\boldsymbol{v}_{\boldsymbol{x}}(\boldsymbol{x},t)
\right)
+
\frac{1}{2}
\nabla_{\boldsymbol{x}}\!\cdot
\left(
\boldsymbol{D}_{\boldsymbol{x}}(\boldsymbol{x},t)
\nabla_{\boldsymbol{x}}\rho(\boldsymbol{x},t)
\right).
\label{eq:FPinx}
\end{equation}
Notably, the dynamics governing $\rho(\boldsymbol{x},t)$ also subscribes to the FP form, albeit a non-canonical version, because the transformed drift velocity $\boldsymbol{v}_{\boldsymbol{x}}(\boldsymbol{x},t)$ is not solely the negative gradient of the pulled-back potential $\psi(T^{-1}(\boldsymbol{x},t))$. The diffusive term, however, retains the canonical form following transformation under the discrepancy map.

The juxtaposition of Equations (\ref{eq:fp_w_pot}) and (\ref{eq:FPinx}) is a physically realized example of the challenge that this work seeks to address: The dynamics of densities in observed space may not adhere to canonical forms (here, of the FP equation) because of diffeomorphic transport maps from a latent space (here the undeformed reference configuration). This insight, rigorously demonstrated for mass transport in deforming solids, could have bearing on successful inference of the transport of probability densities in abstract, high-dimensional settings.

\subsection{Relation to existing approaches}

\Cref{tab:method_comparison} summarizes the relationship between the proposed framework and representative approaches for learning latent dynamics. Existing methods generally fall into two categories.
\begin{itemize}
    \item Transport-based methods, including continuous normalizing flows, flow matching, and dynamic optimal transport, learn time-dependent transport maps while assuming a fixed reference distribution~\cite{Chen2018NeuralODE,
    Lipman2023FlowMatching,BenamouBrenier2000}.
    \item Latent state-space models, including Deep Kalman filters and latent stochastic differential equations, learn evolving latent trajectories together with an observation model~\cite{Kalman1960,Rauch1965,krishnan2015deepkalmanfilters,Li2020LatentSDE}.
\end{itemize}
In contrast, the proposed framework operates directly at the level of evolving probability measures: the latent evolution is constrained by a physically motivated FP equation, while the transport map is learned to account for the remaining geometric discrepancy between latent and observed distributions.

The distinction is particularly important when temporal correspondence between observations is unavailable. In that setting, the primary inference target is not a latent trajectory, but an evolving probability measure whose pushforward through a time-dependent transport map reproduces the observed snapshot distributions. The proposed framework therefore occupies the intersection of these two literatures, but it is not reducible to either one.

\begin{table}[t]
\centering
\fontsize{10}{12}\selectfont
\caption{Conceptual comparison with representative approaches for learning evolving probability distributions.}
\label{tab:method_comparison}
\renewcommand{\arraystretch}{1.15}

\begin{tabular}{p{0.23\textwidth}p{0.22\textwidth}p{0.24\textwidth}p{0.22\textwidth}}
\hline
Method &
Primary object &
Latent evolution &
Observation model
\\
\hline

Continuous normalizing flow
&
Probability distribution
&
Fixed reference distribution
&
Learned transport
\\

Flow matching
&
Probability distribution
&
Fixed reference distribution
&
Learned transport
\\

Dynamic optimal transport
&
Probability distribution
&
Fixed reference distribution
&
Optimal transport map
\\

Deep Kalman filter
&
Latent trajectory
&
Learned stochastic dynamics
&
Learned decoder
\\

Latent SDE
&
Latent trajectory
&
Learned stochastic dynamics
&
Learned decoder
\\

Proposed framework
&
Latent probability measure
&
FP evolution
&
Energy-regularized transport
\\

\hline
\end{tabular}
\end{table}

\subsection{Limitations and future work}

The present work should be viewed as a first step toward a broader framework for population-level dynamical inference over evolving probability measures.

The primary limitation lies in the latent stochastic model. In the present formulation, the latent stochastic evolution is restricted to the OU process. This choice provides an analytically tractable solution of the FP equation and enables efficient optimization, but it also confines the latent probability distributions to the Gaussian family throughout the evolution. Consequently, highly nonlinear or multimodal latent dynamics must be represented primarily through the transport map.

An important direction for future work is therefore the development of richer latent stochastic models that retain analytical or semi-analytical tractability. Possible extensions include Gaussian-mixture models, finite-dimensional invariant families of the FP equation, and more general stochastic processes whose probability evolution admits low-dimensional parameterizations. Such extensions would increase the expressive power of the latent dynamics while preserving the computational advantages of the present framework.

A second limitation concerns the transport representation. The KR rearrangement guarantees invertibility and efficient likelihood evaluation, but it also imposes a triangular structure and an implicit coordinate ordering. Alternative transport parameterizations or adaptive ordering strategies may further improve flexibility in higher-dimensional applications.

From a theoretical perspective, the present work raises several open mathematical questions. The decomposition
$
\rho_t = T_{\sharp}(\cdot,t)\zeta_t
$
is generally non-unique, suggesting connections with inverse problems and gauge symmetries in measure-valued dynamical systems. Establishing conditions for identifiability, existence of minimizers, uniqueness (or uniqueness modulo suitable equivalence classes), and stability of the decomposition remains an important direction for future research. While such questions have been widely explored in nonlinear elasticity, their extension to higher-dimensional probability transport may require further analytic development.

Finally, the current framework assumes complete snapshot observations of the evolving probability distributions. Extending the methodology to irregular sampling, partially observed distributions, higher-dimensional systems, and experimentally measured population data will be essential for applications in computational mechanics, materials science, and computational biology.
\section{Conclusion}
\label{sec:conclusion}
This work introduced a framework for inferring latent stochastic dynamics directly from time-indexed probability distributions without requiring trajectory correspondence. By decomposing the observed probability evolution into intrinsic latent stochastic dynamics and a discrepancy transport map, the proposed approach separates stochastic evolution from geometric deformation, enabling population-level system identification from snapshot observations.

The latent dynamics are represented by an OU process, yielding an analytically tractable FP solution, while the discrepancy transport map is constructed using the KR rearrangement and monotone neural networks. To address the inherent ambiguity of the decomposition, a hyperelastic deformation-energy regularization is incorporated to encourage smooth and physically meaningful transport maps while reducing non-uniqueness. The latent dynamics and transport map are estimated simultaneously through a unified optimization framework over probability distributions.

The numerical examples demonstrate that the proposed framework accurately captures nonlinear and multimodal probability evolution while maintaining a simple latent stochastic representation. These results suggest that combining latent stochastic modeling with transport-based geometric correction provides an effective and interpretable approach for learning distributional dynamics from snapshot data.

Future work will investigate richer classes of latent stochastic processes beyond the OU model, establish theoretical conditions for the existence and identifiability of the latent--transport decomposition, and extend the framework to high-dimensional problems and experimental datasets where only population-level observations are available.

\bibliography{references}
\bibliographystyle{elsarticle-num-names}

\appendix

\end{document}